# Phase Coexistence and the Structure of the Morphotropic Phase Boundary Region in (1-x)Bi(Mg$_{1/2}$Zr$_{1/2}$)O$_3$-xPbTiO$_3$ Piezoceramics

Rishikesh Pandey, Ashish Tiwari, Ashutosh Upadhyay and Akhilesh Kumar Singh*

School of Materials Science & Technology, Indian Institute of Technology (Banaras Hindu University) Varanasi- 221005, India

## Abstract

Structure of the morphotropic phase and the phase coexistence region has been investigated in (1-x)Bi(Mg$_{1/2}$Zr$_{1/2}$)O$_3$-xPbTiO$_3$ ceramics. The structure is cubic with space group Pm3m for the compositions with x<0.57 and tetragonal with space group P4mm for the compositions with x>0.59. For the compositions with 0.56<x<0.60, both the tetragonal and cubic phases coexist, which suggest very narrow morphotropic phase boundary region of compositional width Δx~0.03. Rietveld refinement of the structure using x-ray diffraction data confirms coexistence of the tetragonal and cubic phases in the MPB region and rules out the coexistence of tetragonal and rhombohedral structure reported by earlier workers. After poling appearance of significant value of electromechanical coupling coefficient in the cubic compositions (x=0.55, 0.56) suggests the presence of electric field induced transition from centrosymmetric cubic phase to noncentrosymmetric phase. Polarization- Electric field hysteresis loop measurement on the cubic composition with x=0.56 gives well saturated loop similar to that observed in Pb(Mg$_{1/3}$Nb$_{2/3}$)O$_3$ relaxor below the freezing temperature.

* Corresponding author

E-mail address: akhilesh_bhu@yahoo.com, aksingh.mst@itbhu.ac.in

## 1. Introduction

From last several decades the Pb-based piezoceramics such as $Pb(Zr_xTi_{1-x})O_3$ (PZT) and $Pb(Mg_{1/3}Nb_{2/3})O_3$-$xPbTiO_3$ (PMN-PT) have been the best choice as active materials for device applications in electromechanical transducers and actuators [1-2]. In recent years, there is great interest in lead free piezoceramics because of worldwide growing concern over toxicity of lead [3-6]. This has shifted the attention of researchers towards Bi-based piezoelectric ceramics and morphotropic phase boundary (MPB) solid solutions [7]. Several new Bi-based MPB systems have been investigated recently and some of them have added advantage of $T_C$ higher than PZT making them more useful for high temperature piezoelectric applications [8-12]. In the search of such high temperature stable perovskites, solid solution of $BiScO_3$ with $PbTiO_3$ i.e. $(1-x)BiScO_3$-$xPbTiO_3$ (BS-PT) has been investigated with high transition temperature ($T_C$~ 460 $^0$C) higher than PZT (~ 400 $^0$C) and tremendous piezoelectric response near MPB around x~0.64 [9]. However, because of costly 'Sc' this system is not cost effective. Several other Bi-based solid solutions such as $(1-x)Bi(Ni_{1/2}Ti_{1/2})O_3$-$xPbTiO_3$ [13], $(1-x)Bi(Ni_{1/2}Hf_{1/2})O_3$-$xPbTiO_3$ [14], $(1-x)Bi(Zn_{1/2}Ti_{1/2})O_3$-$xPbTiO_3$ [15], $(1-x)Bi(Mg_{1/2}Ti_{1/2})O_3$-$xPbTiO_3$ (BMT-PT) [8], $(Bi_{0.5}Na_{0.5})TiO_3$-$BaTiO_3$-$(Bi_{0.5}K_{0.5})TiO_3$ (BNT-BT-BKT), and $(1-x)Bi(Mg_{1/2}Zr_{1/2})O_3$-$xPbTiO_3$ (BMZ-PT) [5-7, 16-17] have been explored and some of them have significantly higher transition temperature ($T_C$) than PZT making them superior for high temperature piezoelectric applications. In BMZ-PT solid solution, Suchomel et al. [7] have reported recently that the structure is cubic at x=0.55 and tetragonal at x=0.60 which indicates that MPB for this system may lie somewhere in between 0.55 to 0.60. Shabbir et al. [16] and Qureshi et al. [17] studied the structure and dielectric properties of this solid solution and reported MPB around ~0.54 to 0.57 and 0.55 to 0.60, respectively. However, the piezoelectric characterization of BMZ-PT has not been done by these workers. Further, the structure of MPB was reported to be the coexistence of rhombohedral and tetragonal structures, but the x-ray diffraction patterns shown by these workers do not support this fact. It is well-known that the crystal structure of the MPB compositions plays very important role in high piezoelectric response at MPB and very small variation (~0.05 in PZT [18]) in composition can change the crystal structure and significantly decrease the piezoelectric response. In view of this we decided to carryout detailed investigation of the structure and piezoelectric response of BMZ-PT solid- solution on the compositions prepared at 1% interval. We find that the width of the MPB region is very narrow ($\Delta x \sim 0.03$) in this solid- solution similar to PZT [18] and $(1-x)BiFeO_3$-$xPbTiO_3$ (BF-PT) [19], but the structure of the MPB composition at room temperature consists of coexisting tetragonal and cubic phases, and not monoclinic as observed in high piezo-response ceramics such as PZT [20] and PMN-PT [21].

**2. Experimental procedure**

$(1-x)Bi(Mg_{1/2}Zr_{1/2})O_3$-$xPbTiO_3$ ceramics with x=0.55, 0.56, 0.57, 0.58, 0.59 and 0.60 were prepared by traditional solid state ceramic route. Analytical reagent grade $Bi_2O_3$, $ZrO_2$, MgO, $TiO_2$, and PbO of purity > 99% obtained from HiMedia Laboratories Pvt. Ltd. were used as raw materials. Stoichiometric amount of these powders with acetone as mixing media were mixed for 6 h in a planetary ball mill (Retsch, Germany) using agate jar and balls. This mixture was dried and then calcined inside muffle furnace for 6 h at optimized temperature of 775 $^0$C. The calcined powders were checked for phase purity using an 18 kW, rotating Cu target, Rigaku (Japan) x-ray diffractometer operating in the Bragg-Brentano geometry with the curved crystal graphite monochromator fitted in the diffracted beam. The instrumental resolution is 0.10 degree. The XRD data were collected at a scan rate of 2 degrees/min in the 2-theta range from $20^0$- $120^0$ at the scan step of $0.02^0$. For pellet formation, 2% polyvinyl alcohol solution in water was used as binder. Cold compaction of calcined powder was done at an optimized load of 65 kN using a steel die of 12 mm diameter and an uniaxial hydraulic press. The green pellets were kept at 500 $^0$C for 10 h to burn off the binder material and then sintered at optimized temperature of 975 $^0$C for 3 h in sealed crucibles with controlled PbO/ Bi$_2$O$_3$ atmosphere using small amount of PbO and Bi$_2$O$_3$ as sacrificial powder. Density of the sintered pellets was higher than 98% of the theoretical density. For recording XRD patterns, sintered pellets were crushed into fine powders and then annealed at 500 $^0$C for 10 h to remove the strains introduced during crushing. The microstructure of the as sintered sample surface was studied by Scanning electron microscope (SEM) using ZEISS SUPRA40. Thin gold film was sputter coated on the sintered pellets before examining under SEM. For electroding, flat surfaces of the sintered pellets were gently polished with 0.25 μm diamond paste and washed with acetone. Isopropyl alcohol was applied to clean the surfaces and removing the moisture, if any. Fired on silver paste was subsequently applied on both the surfaces of the pellet. It was first dried around 120 $^0$C in an oven and then cured by firing at 500 $^0$C for 5 minutes. The electroded pellets were poled in silicon oil bath cooling from 100 $^0$C to room temperature under 30kV/cm DC field. Electromechanical coupling coefficients of poled pellets were measured by resonance- antiresonance method [1]. For polarization (P-E) measurements, Radiant Ferroelectric Loop Tracer (USA) was used. The structural analysis by Rietveld method was done using FullProf Suite [22]. In all structural refinements origin was fixed at A-site cations (Pb$^{2+}$/Bi$^{3+}$) except for the rhombohedral structure. To fix the origin for the rhombohedral structure, the z-coordinate of O$^{2-}$ ion was fixed at 1/6. Pseudo-Voigt function was used to model the XRD profiles while anisotropic peak broadening functions suggested by Stephens [23] was used to fit the anisotropic broadening of reflections. Background was modelled by 5$^{th}$ order polynomial.

**3. Results and discussion**

*3.1. Crystal structure: A Rietveld study*

Powder XRD patterns of BMZ-PT ceramics with x=0.55, 0.56, 0.57, 0.58, 0.59 and 0.60 collected at room temperature for the 2θ range of 20 to 60 degrees is shown in Fig.1. All the reflections shown in Fig.1 correspond to perovskite structure except negligibly weak reflections (marked with asterisks) around 2-theta ≈ $27.58^0$, $28.69^0$ and $32.82^0$ in the XRD profile. Similar impurity reflections are also seen in the XRD patterns of BMZ-PT samples reported by earlier workers [16-17]. To eliminate these impurity phases one needs multiple calcination and sintering steps [7]. Our samples were prepared in single step calcination and sintering schedule. The phase fraction for this impurity phase is negligibly small as determine by using Rietveld method. A careful examination of Fig.1 suggests the presence of three different crystallographic compositional regions, (i) x>0.59 (ii) x<0.57 and (iii) 0.57≤ x ≤0.59, where the nature of the XRD profiles changes (see e.g. the peak around 2θ= $46^0$ encircled in Fig.1). To analyze the structure, we have shown in Fig.2, the selected pseudocubic (110), (111) and (200) profiles of BMZ-PT ceramics with x=0.55, 0.56, 0.57, 0.58, 0.59 and 0.60. In all the profiles shown in this figure, the contribution from $Cu_{K\alpha2}$ wavelength has been subtracted by using standard software. For the compositions with x≤0.56, all the pseudocubic reflections shown in Fig.2 appear to be singlet. This characterizes a cubic structure with space group Pm3m for the compositions with x≤0.56. This is consistent with the results of Suchomel et al. [7], where cubic structure is reported for x=0.55. For the composition with x=0.60, (110) and (200) profiles are clearly splitted into doublet while (111) profile remains singlet. This characterises tetragonal structure for x≥0.60. Earlier workers have also reported tetragonal structure with space group P4mm for x=0.60 [7, 16]. For the compositions with x=0.57, 0.58 and 0.59, the examination of (200) pseudocubic XRD profile shown in Fig.2 suggest that the peaks corresponding to both the cubic and tetragonal phases are present indicating coexistence of these two phases. Thus the compositions with 0.57≤x≤0.59 correspond to the MPB region in which both the tetragonal and cubic phases coexist. As shown in Fig.2, the intensity of the peak corresponding to cubic phase marked by 'C' is dominant in x=0.57 and 0.58 while the intensity of the peak corresponding to tetragonal phase marked by 'T' is dominant in x=0.59. There is no lower angle splitting or asymmetry in (111) profile expected for the rhombohedral/monoclinic structures. This suggests that the rhombohedral structure is not present in or outside the MPB region in BMZ-PT ceramics.

To confirm the coexistence of tetragonal and cubic phases and for the sake of completeness we have analyzed the crystal structure of MPB compositions [$0.57 \leq x \leq 0.59$, $\Delta x = 0.03$] using various plausible structures i.e. rhombohedral (R3m), cubic (Pm3m) and monoclinic (Pm and Cm) coexisting with the tetragonal (P4mm) structure. Our Rietveld analysis of the XRD data for these compositions reveals that the best fit is obtained for the coexistence of cubic and tetragonal (Pm3m+P4mm) phases. The value of $\chi^2 = 2.19$ is also lowest for this structural model as against the (R3m+P4mm) model giving $\chi^2 = 3.80$ for x=0.59. Thus we may conclude that the structure for the compositions with $0.57 \leq x \leq 0.59$ consists of coexisting cubic and tetragonal phases (Pm3m+P4mm) and not the rhombohedral and tetragonal (R3m+P4mm) phases reported by earlier workers [16-17]. This is further, confirmed by the absence of splitting in the (111) pseudocubic XRD profiles expected for the rhombohedral structure (see Fig.2 and inset to Fig.3(c)). In contrast, the MPB systems such as PZT [20] and PMN-PT [21] show clear splitting in (111) pseudocubic XRD profile for the compositions with rhombohedral structure. First-principles density functional theory calculations by Grinberg and Rappe [24] also suggest that $T_C$ decreases rapidly for BMZ-PT solid solution with decreasing PT concentration and approaches to 300K for $x \approx 0.50$. The Rietveld structure refinement of all the compositions has been done to determine precisely the structural parameters. The Rietveld fits for x=0.56, 0.59 and x=0.60 using cubic, cubic+ tetragonal and tetragonal structures respectively, are shown in Fig.3. The Rietveld analysis also confirms the tetragonal structure for x=0.60. Very good fit between the observed and calculated profiles were obtained using tetragonal P4mm space group [see Fig.3(a) for x=0.60]. Consideration of small coexisting cubic phase for x=0.60 further improve the fit however, the structure is predominantly tetragonal. Fig.3(c) depicts the Rietveld fit of the XRD data for x=0.56 using cubic space group Pm3m. The fit between the observed and calculated profiles is quite good confirming the cubic structure of BMZ-PT for x<0.57. As shown in Fig. 3(b) the Rietveld fit for the composition with x=0.59 using coexistence of cubic and tetragonal (Pm3m+P4mm) phases is also quite good. A similar coexistence of the cubic and tetragonal phases in the MPB region has been reported in the (1-x)Pb($Fe_{2/3}W_{1/3}$)$O_3$-xPbTiO$_3$ ceramics [25], but with very wide phase coexistence region (0.20≤x≤0.37). The compositional width of the MPB is very narrow in the BMZ-PT system similar to PZT [18]. The fact that some MPB systems like PZT [20], BS-PT [9-12], BF-PT [19], BMZ-PT etc. have very narrow compositional width of the phase coexistence region while the MPB systems such as PMN-PT [21], BMT-PT [8], (1-x)Pb($Fe_{2/3}W_{1/3}$)$O_3$-xPbTiO$_3$ [25] etc. have wide phase coexistence region is very much intriguing. Suchomel et al. [7] have successfully investigated the possible relationship between the tolerance factor and the position and compositional width of the phase coexistence region for MPB ceramics. It has been reported that the position of MPB shifts to the PbTiO$_3$ rich end as the difference of the tolerance factor with the other end component increases. However, increase in the difference in tolerance factor results in narrow compositional width of the MPB. This is indeed the case for the MPB ceramics such as PZT [20], BS-PT [9-12] as well as for the BMZ-PT system. The BF-PT [19] may be an exception for which the MPB lies on the lower PT side, while the compositional width of the MPB is also narrow. The presence of rotational instabilities in the rhombohedral structure of BiFeO$_3$ may be the possible reason for this anomaly [19]. More detailed theoretical studies will be needed to understand the location and the compositional width of the MPB in 'Pb' and Bi-based MPB systems. Further, in PMN-PT system the structure of the component PMN is pseudocubic but in the vicinity of the MPB short range monoclinic followed by long range monoclinic structure is stabilized with $PbTiO_3$ substitution [21]. Theoretical studies will be needed to understand why similar rhombohedral phase is not stabilized in the vicinity of the MPB in BMZ-PT system.

The refined structural parameters are listed in Table.1 along with the agreement factors for the compositions with x=0.59 and 0.60. As observed in several Pb-based piezoelectric ceramics [21], the isotropic thermal parameters for the A-site cations ($Pb^{2+}/Bi^{3+}$) were very high (>4 $Å^2$). In view of this we considered anisotropic thermal parameters for these ions in tetragonal structure. As can be seen in table I, the thermal ellipsoid is more elongated about xy-axes than the z-axis. For cubic phase of the x=0.59 composition, the refined thermal parameters for oxygen and B-site cations ($Mg^{2+}/Zr^{4+}/Ti^{4+}$) were also very high. However, fixing their values to low level manually do not affect significantly the value of $\chi^2$ suggesting these parameters are not correctly determined with the XRD data. Powder neutron diffraction data will be needed to ascertain correctly the thermal parameters for these ions.

Variation of the tetragonal phase fraction and lattice parameters with composition, as obtained by Rietveld refinement of the structure are plotted in Fig.4. It is evident from this figure that very small change in PT concentration (x) affects significantly the fraction of tetragonal phase. As expected, the tetragonality increases with increasing PT concentration for tetragonal phase region. The values of the lattice parameters for x= 0.55 and 0.60 are in good agreement with that reported by Suchomel et al. [7].

3.2. Microstructure Studies

SEM micrographs of (1-x)BMZ-xPT ceramics with composition x=0.56, 0.58 and 0.60 are shown in Fig.5(a), Fig.5(b) and Fig.5(c) respectively. The average grain size for

x=0.56, 0.58 and 0.60 is obtained to be ~ 5.34 μm, 6.54 μm and 6.75 μm respectively, which suggests slightly increasing trend in the grain size with increasing PT concentration. Since $Bi_2O_3$ is well known grain size refiner, smaller grain size is expected in the Bi-rich compositions. There are few dark grains at the junction of the grain boundaries marked by the arrow in Fig.5(a) and Fig.5(b). These dark grains are due to an impurity phase as confirmed by the energy dispersive spectroscopy (EDS) studies. Presence of small impurity phase is also evident in the XRD studies. The EDS spectra of the perovskite and impurity phases for x=0.56 are shown in Fig.6 (a) and Fig.6(b) respectively. A comparison of Fig.6 (a) and (b) suggests that the impurity phase is significantly deficient in Ti and also Pb and Bi. Quantitative analysis of the impurity phase reveals that it is mostly Mg and O based compound with small concentration of Pb and Bi.

3.3 *Polarization Studies*

Polarization- Electric field (P-E) hysteresis loops for the compositions with x=0.56 (cubic), 0.58 (cubic+Tetragonal) and x=0.60 (tetragonal) are shown in Fig.7(a), 7(b) and 7(c) respectively. As can be seen from Fig.7(a) a well saturated hysteresis loop is observed for x=0.56. Since the structure of this composition is cubic at room temperature one should not get the hysteresis loop if it is the centrosymmetric paraelectric phase. This anomalous result could be explained only if we consider the presence of the relaxor behaviour in this composition. In fact, temperature dependence of the dielectric permittivity of BMZ-PT with x=0.55 reported by Suchomel [7] et al. exhibit a strong peak around ~ 250 $^0$C ($T_m$) with large frequency dispersion suggesting presence of strong relaxor features. In canonical relaxor PMN it is known that the average structure remains cubic down to 4K temperature [26] because of the fact that the development of long range ferroelectric order is prevented by the presence of the quenched random fields [27] resulting from the presence of off-valent $Mg^{2+}$/$Nb^{5+}$ ions at the B-site. However, temperature dependence of the dielectric permittivity exhibit strong peak with huge dielectric permittivity and frequency dispersion around 290K. Recently in case of Pb(Mg$_{1/3}$Nb$_{2/3}$)O$_3$ (PMN) [28] also, it has been reported that a well saturated hysteresis loop similar to normal ferroelectrics is observed when temperature is lowered below freezing temperature (T$_{vf}$=220K) [29-30]. Since T$_m$=250 $^0$C [7] for BMZ-PT with x=0.55, the freezing temperature can be well above the room temperature resulting in the well saturated hysteresis loop in average cubic symmetry similar to PMN. Appearance of well saturated hysteresis loop for x=0.55 misled to Shabbir et al [16] and Qureshi et al [17] to draw the conclusion that this compositions is rhombohedral even though no splitting were seen in the XRD profiles. Surprisingly, the hysteresis loops shown in Fig.7(c) for the tetragonal phase with x=0.60, is very slim in comparison to that for x=0.55 and saturation is not reached. Similarly the hysteresis loop for the composition with x=0.58 showing the coexistence of the cubic and tetragonal phases is also less saturated than that observed for x=0.55. It seems that the reorientation of the nanodomains in x=0.55 is much easier than the microscopic domains in x=0.60 preventing the saturation of the hysteresis loop. We tried to open the loop at higher electric fields but dielectric breakdown occurred. Detailed studies of these aspects are underway and will be reported elsewhere.

*3.4 Composition dependence of the Electromechanical Coupling Coefficient (k$_P$)*

Fig.8 shows the variation of planer electromechanical coupling coefficient (k$_P$) with composition for BMZ-PT ceramics. The value of k$_P$ is significantly lower (~0.26) than the widely used MPB systems such as PZT and PMN-PT. As reported by Fu and Cohen [31], the high piezoelectric response near MPB is linked with the ease of rotation of the polarization vector under external stimuli like electric field or stress. The presence of monoclinic phase in the MPB region provides a low energy path way for the rotation of the polarization vector. This is the reason why the Bi-based MPB systems like BS-PT, also exhibit high piezoelectric response for the MPB compositions with monoclinic structure [9-12]. The presence of cubic phase coexisting with the tetragonal phase in the BMZ-PT system may be attributed for lower $k_P$ as against PZT [2, 18], PMN-PT [2, 21] and BS-PT [9-12]. Further, as shown in Fig.5, appreciable electromechanical coupling coefficient $k_P$ (~0.21) is obtained for the cubic composition with x=0.55 also, which is not expected for the centrosymmetric structure. This could be explained only if there is an electric field induced phase transition from paraelectric cubic phase to a ferroelectric phase with non-centrosymmetric structure. Electric field induced phase transition from pseudocubic phase to a ferroelectric phase have been observed recently in pseudocubic compositions of several Bi-based solid solutions [32-34]. Possibly a similar phase transition is taking place in x=0.55 also during poling of the samples. Temperature dependence of the dielectric permittivity of BMZ-PT with x=0.55 reported by Suchomel [7] et al. exhibit a strong peak around ~250 $^0$C with large frequency dispersion which is not expected for the paraelectric cubic structure. This suggests that the cubic structure of x=0.55 composition is because of the strong relaxor features similar to that reported in canonical relaxor PMN [35-36]. It has been reported for PMN that a field induced relaxor to ferroelectric phase transition takes place under the application of electric field [37-38]. A similar phase transition from pseudocubic relaxor to noncentrosymmetric phase in 0.45BMZ-0.55PT can be attributed to the appearance of piezoelectric response after poling. It will be interesting to study the depoling temperature and temperature of transition to relaxor/paraelectric phase, for this ferroelectric phase, resulting from field induced transition. Appearance of piezoelectric response in pseudocubic compositions of BMZ-PT suggests that apart from the non-centrosymmetric structures of solid solutions, the pseudocubic compositions should also be explored for the piezoelectric properties.

## 4. Conclusions

To conclude, detailed Rietveld analysis of the structure of BMZ-PT for the compositions with 0.55≤x≤0.60 reveals that the structure is cubic in the Pm3m space group for the compositions with x<0.57, tetragonal in the space group P4mm for the compositions with x>0.59 while the two phases coexists for the intermediate compositions. Thus the width of the MPB region is very narrow Δx~0.03. Our results clearly rule out the coexistence of tetragonal and rhombohedral structure for the MPB compositions reported by earlier workers. Highest electromechanical coupling coefficient obtained for the composition with x=0.57 is 0.26, which is significantly lower than PZT, because of the fact that there is no monoclinic phase in the MPB region of BMZ-PT. After poling the cubic compositions (x=0.55, 0.56) also show piezoelectricity suggesting an electric field induced phase transition to a ferroelectric phase. The BMZ-PT Composition with x=0.56 having cubic structure shows well saturated hysteresis loop similar to that observed for well known relaxor PMN below freezing temperature. Our results would further encourage *ab-initio* first principles calculations to understand the nature of the MPB in the Bi-based ceramics and help in developing lead free piezo-ceramics with higher piezoelectric response.

**Acknowledgement**

RP acknowledges University Grant Commission (UGC), India for the financial support as Senior Research Fellowship (SRF).

**FIG.1**. Powder XRD patterns of $(1-x)Bi(Mg_{1/2}Zr_{1/2})O_3$-$xPbTiO_3$ ceramics with x=0.55, 0.56, 0.57, 0.58, 0.59 and 0.60 sintered at 975 $^0$C. Weak reflections marked by asterisk are due to impurity phase. Miller indices are given for tetragonal structure with x=0.60.

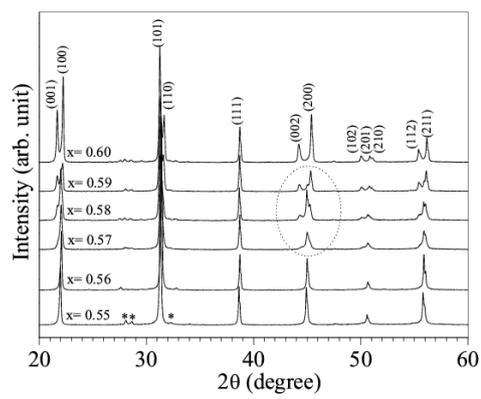

Fig.1

**FIG.2**. Pseudocubic (110), (111) and (200) powder XRD profiles of $(1-x)Bi(Mg_{1/2}Zr_{1/2})O_3$-$xPbTiO_3$ ceramics for the compositions with x=0.55, 0.56, 0.57, 0.58, 0.59 and 0.60. The peaks marked with 'C' and 'T' correspond to cubic and tetragonal phases respectively.

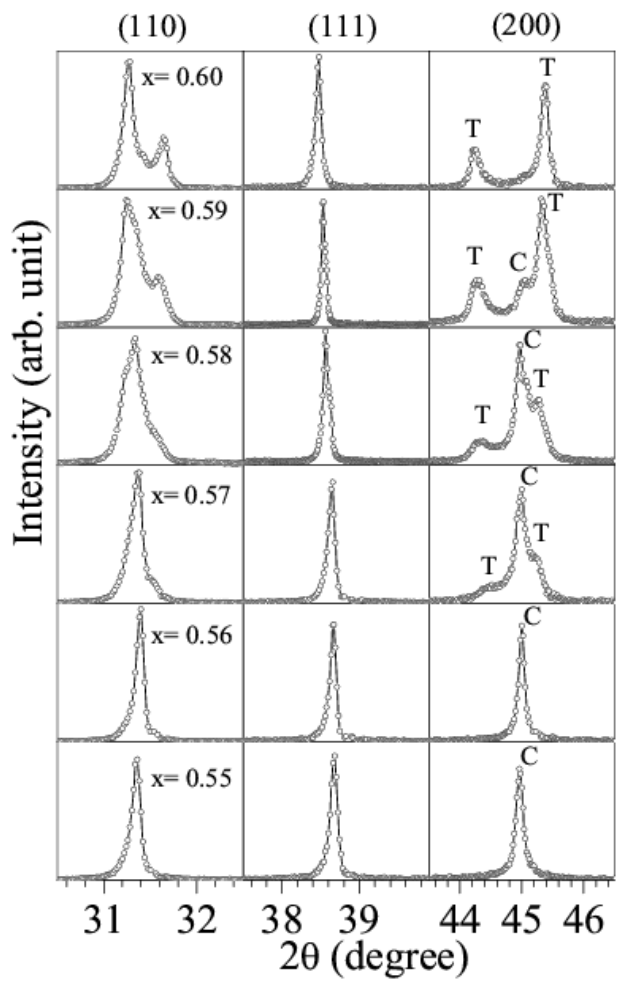

Fig.2

**FIG.3**. Observed (dots), calculated (continuous line) and difference (continuous bottom line) profiles for (1-x)Bi(Mg$_{1/2}$Zr$_{1/2}$)O$_3$-xPbTiO$_3$ with (a) x=0.60, (b) x=0.59 and (c) x=0.56 obtained after Rietveld analysis of the powder XRD data using tetragonal (P4mm), tetragonal+cubic (Pm3m+P4mm) and cubic (Pm3m) structures respectively. The vertical tick marks above the difference plot show the peak positions. The insets illustrate the goodness of fit.

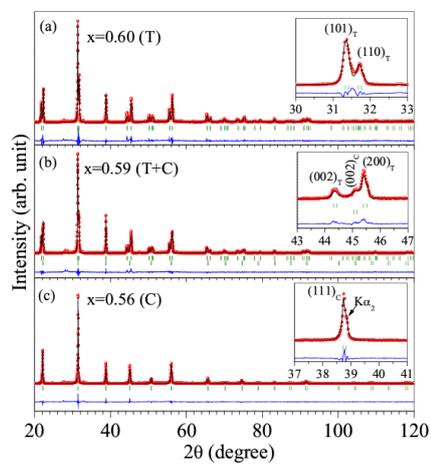

Fig.3

**FIG.4**. Compositional variation of cubic ($a_C$), tetragonal ($a_T$, $c_T$) lattice parameters and tetragonal Phase fraction ($P_T$) for $(1-x)Bi(Mg_{1/2}Zr_{1/2})O_3$-$xPbTiO_3$ ceramics obtained after Rietveld analysis. Continuous lines are guide for eyes.

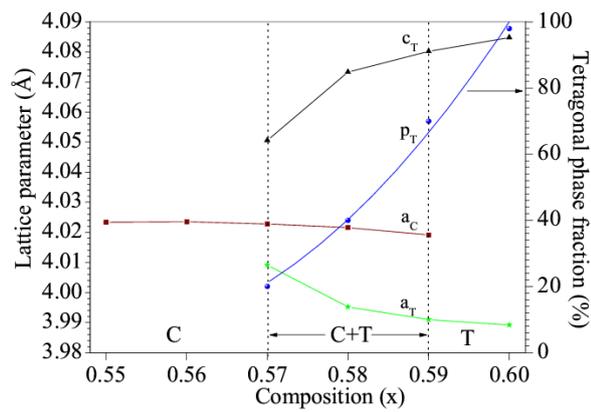

Fig.4

**FIG.5**. SEM images of (1-x)Bi(Mg$_{1/2}$Zr$_{1/2}$)O$_3$-xPbTiO$_3$ ceramics for (a) x=0.56 (b) x=0.58 and (c) x=0.60. The dark grains marked by arrow correspond to impurity phase.

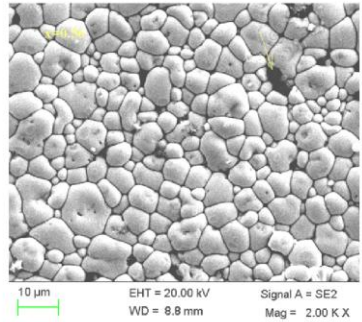

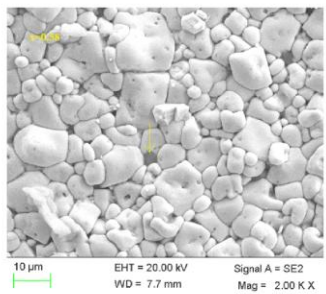

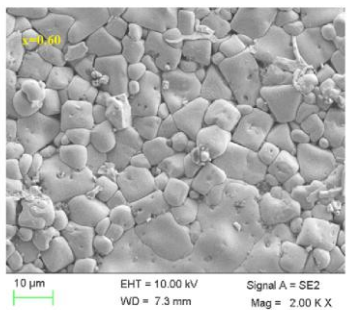

**FIG.6**. EDS spectrum of the (a) perovskite phase and (b) impurity phase of the (1-x)Bi(Mg$_{1/2}$Zr$_{1/2}$)O$_3$-xPbTiO$_3$ ceramics for the composition with x=0.56.

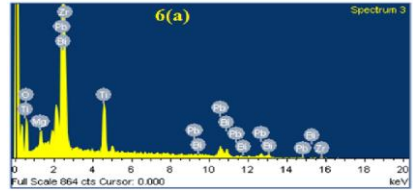

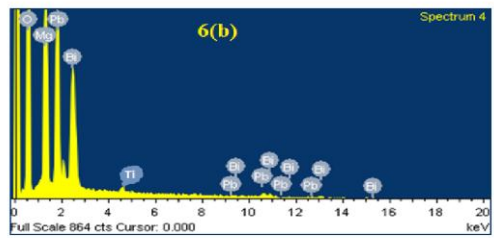

**FIG.7**. P-E hysteresis loop for the $(1-x)Bi(Mg_{1/2}Zr_{1/2})O_3$-$xPbTiO_3$ ceramics with (a) x=0.56, (b) x=0.58 and (c) x=0.60 recorded at 50Hz.

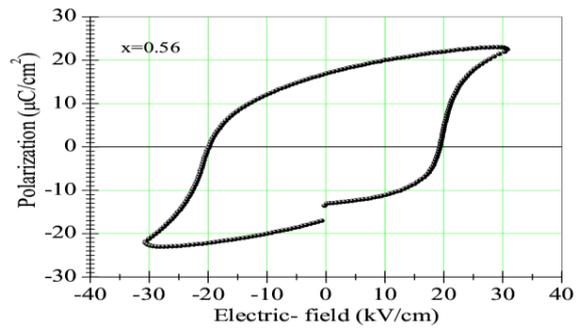

Fig7(a)

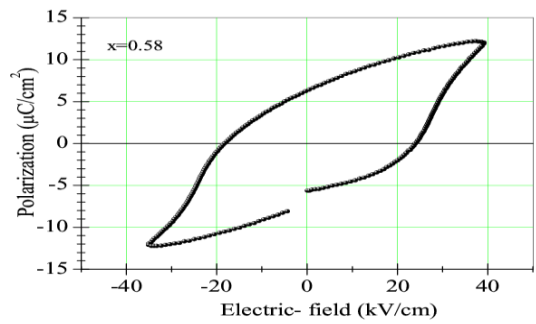

Fig7(b)

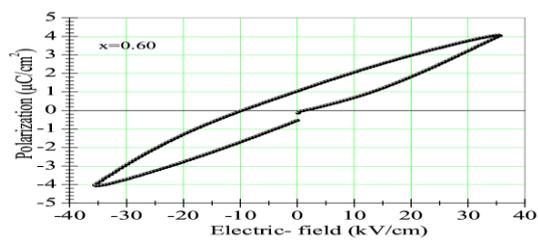

Fig7(c)

**FIG.8**. Variation of electromechanical coupling coefficient ($k_P$) with composition for $(1-x)Bi(Mg_{1/2}Zr_{1/2})O_3$-$xPbTiO_3$ ceramics. Continuous lines are guide for eyes.

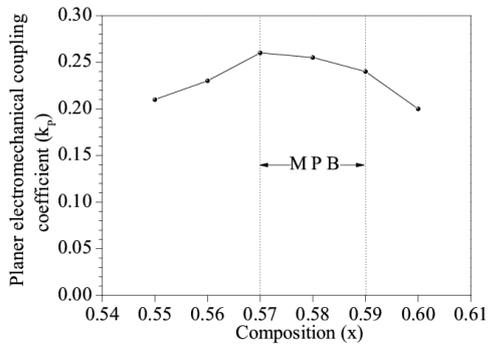

Fig.8